\documentclass[12pt,preprint]{aastex}

\newcommand{\e}            {\mbox{$^{-1}$}}

\newcommand{\Mearth}       {\mbox{$M_\oplus$}}
\newcommand{\persqcm}      {\rm \,cm^{-2}}
\newcommand{\percc}        {\rm \,cm^{-3}}
\newcommand{\kms}          {km~s$^{-1}$}
\newcommand{\pp}           {\noindent\hangindent 20pt\hangafter=1}

\def\plotfiddle#1#2#3#4#5#6#7{\centering \leavevmode
\vbox to#2{\rule{0pt}{#2}}
\includegraphics{#1}}

\begin{document}

\title{An 850~\micron\ survey for dust around solar mass stars}
\author{Joan Najita}
\affil{National Optical Astronomy Observatory, 950 North Cherry Avenue,
Tucson, AZ 85719}
\email{najita@noao.edu}
\author{Jonathan P. Williams}
\affil{Institute for Astronomy, 2680 Woodlawn Drive, Honolulu, HI 96822}
\email{jpw@ifa.hawaii.edu}

\shorttitle{Survey for dust around solar mass stars}
\shortauthors{Najita \& Williams}

\begin{abstract}
We present the results of an 850~\micron\ JCMT/SCUBA survey for dust
around 13 nearby solar mass stars.
The dust mass sensitivity ranged from $5\times 10^{-3}$ to $0.16~M_\oplus$.
Three sources were detected in the survey, one of which (HD~107146) 
has been previously reported. 
One of the other two submillimeter sources, HD~104860, was not detected
by IRAS and is surrounded by a cold, massive dust disk with a dust 
temperature and mass of $T_{dust}=33$~K and $M_{dust}=0.16~M_\oplus$.
The third source, HD~8907, was detected by IRAS and ISO at $60-87$~\micron,
and has a dust temperature and mass 
of $T_{dust}=48$~K and $M_{dust}=0.036~M_\oplus$.
We find that the deduced masses and radii of the dust disks in our sample 
are roughly consistent with models for the collisional evolution 
of planetesimal disks with embedded planets.   
We also searched for residual gas in two of the three systems 
with detected submillimeter excesses and place limits on the mass of 
gas residing in these systems.  

When the properties measured for the detected excess sources are
combined with the larger population of submillimeter excess sources
from the literature, we find strong evidence that the mass in small
grains declines significantly on a $\sim 200$ Myr timescale,
approximately inversely with age.
However, we also find that the characteristic dust radii of the 
population, obtained from the dust temperature of the excess and 
assuming blackbody grains, is uncorrelated with age. 
This is in contrast to self-stirred collisional models for debris 
disk evolution which predict a trend of radius increasing with 
age $t_{age} \propto R_d^3$. 
The lack of agreement suggests that processes beyond self-stirring, 
such as giant planet formation, play a role in the evolutionary 
histories of planetesimal disks. 
\end{abstract}

\section{Introduction}

The existence of dust around stars well beyond the protostellar
phase is thought to arise from the collisions of planetesimals, the
building blocks of planets, and is therefore of great interest for
the clues they may reveal about the processes and timescales involved 
in planet formation (Backman \& Paresce 1993).  Indeed dusty disks
are thought to be one of the most readily observable signatures of
extrasolar planetary systems.  Abundant dust is expected to be
produced early on as part of the planet formation process itself,
as planetesimals collide and grow into planetary mass objects (e.g.,
Kenyon \& Bromley 2004a, b).  Further dust production is expected as
residual planetesimal disks grind themselves down to the low masses
inferred for the Kuiper Belt today (e.g., Farinella, David, \&
Stern 2000).  Giant planets in the system may sculpt the resulting 
debris, providing an indirect signature of their presence 
(e.g., Liou \& Zook 1999; Ozernoy et al.\ 2000; Quillen \& Thorndike 2002; 
Moro-Mart\'in \& Malhotra 2002). 
In the context of our solar system, the age range
10--100 Myr is of interest, since it overlaps the epoch of giant planet 
formation and includes the epoch of formation of large bodies in the 
outer solar system.  Also during this time period, the interaction
of giant planets with a residual planetesimal disk may lead to
planetary orbital migration and sculpting of the planetesimal disk.

Space based far-infrared telescopes have discovered and begun to
characterize the properties of dusty disks around main sequence stars
(Aumann 1985, Spangler 2000)
and it is one of the primary goals of the recently launched
{\it Spitzer Space Telescope}.
Ground-based sub-millimeter observations provide 
useful complementary
information about the properties of such disks. First, the dust mass
is readily measured because the emission is almost certainly optically
thin at long wavelengths.
Second, by anchoring the 
Rayleigh-Jeans side of the spectral energy
distribution (SED), contraints can be placed on the temperature and
size of the emitting grains.
Finally, the relatively high resolution that can be attained at these
wavelengths allows
investigations of the morphology of the emitting dust.  
Dust morphologies have been used to argue for the presence of 
perturbing planetary companions 
(Holland et al.\ 1998, Greaves et al.\ 1998, Wilner et al.\ 2002).

In this paper we present the results of a sub-millimeter search 
for dust debris surrounding 13 solar mass stars that are targets of 
the {\it Spitzer} Legacy Science Project, ``Formation and Evolution
of Planetary Systems'' (FEPS; Meyer et al. 2004).
Since a goal of the FEPS project is to place our solar system in
context, the parent sample covers spectral types corresponding
to approximately solar mass stars spanning a range of ages
from 3 Myr to 3 Gyr.
The submillimeter targets were selected from the FEPS sample with
an emphasis on sources that are young (ages 10 to a few 100 Myr) and
nearby ($\le 50$ pc) based on estimated stellar ages for the FEPS sample
(Hillenbrand et al.\ 2005) and Hipparcos distances.
We emphasized nearby sources
in order to enable the detection of faint excesses or to
place meaningful upper limits on the submillimeter flux.
Our sample is shown in Table 1.

\section{Observations}
Photometry observations were conducted using the SCUBA bolometer
at the James Clerk Maxwell Telescope (JCMT) on Mauna Kea, Hawaii
between February 2003 and January 2004.
The survey was only carried out in stable weather conditions when
the precipitable water vapor level was less than 2~mm.
Zenith optical depths ranged from 0.1 to 0.2 at 850~\micron\ 
and 0.3 to 0.8 at 450~\micron\ and the integration time spent on
each source was between 40 minutes to 3 hours. The resulting
rms errors varied from 1 to 3~mJy~beam\e\ at 850~\micron\
and 10 to 30~mJy~beam\e\ at 450~\micron.
The pointing was checked via observations of bright quasars near
each source after each major slew and the focus was adjusted
every three hours on average, more frequently at times near
sunrise and sunset. Calibration was performed by observations
of Uranus, Neptune, and Mars and standard sources,
CRL 618, CRL 2688, IRC+10216. Based on the agreement of the measured
fluxes with those predicted for these calibrators, we estimate
the photometry is accurate to within 10\% at 850~\micron\
and 30\% at 450~\micron.

Additional CO(3--2) observations of HD~107146 and HD~104860
were made using the 345 GHz receiver and 500~MHz
acousto-optical spectrometer at the Caltech Submillimeter Observatory
(CSO) in March 2003. The beamsize at this frequency is $15''$.
The weather was very dry and stable resulting in system temperature that
ranged from 450~K to 550~K.
Spectra were taken in chopping mode with a throw of $120''$ at 0.321~Hz.
The passbands were very flat over the frequency interval of interest
and a single order baseline was removed
from individual spectra before coadding and rebinning to 1~\kms\ channels.
Pointing was checked via observations of IRC+10216 and was accurate to $2''$.
Antenna temperatures were converted to the main beam scale using an
efficiency $\eta_{\rm mb}=0.65$.
No emission was detected in either source at an rms noise level of
5~mK (HD~107146) and 13~mK (HD~104860) per channel.

\section{Results}
\subsection{Submillimeter Continuum} 
We detected 3 objects out of our sample of 13. The measured fluxes and 
upper limits for the entire sample are shown in Table 1.
The errors and upper limits in this Table are based on the
rms noise only and do not include the calibration uncertainty
discussed above.
The results are also shown graphically in the spectral energy
distributions (SEDs) plotted in Figures 1, 2, and 3.
To construct the SEDs, we tabulated infrared fluxes from the
2MASS All-Sky Survey, the IRAS Faint Source Catalog (FSC),
and additional ISO and submillimeter fluxes from the literature 
(Silverstone 2000; Carpenter et al.\ 2005; Liu et al.\ 2004; 
Habing et al.\ 2001).
IRAS detections were color corrected based on the effective temperature
of the star for 12 and 25~\micron\ and (typically) for a 50~K disk at 
60 and 100~\micron (cf. HD104860, Figure 1, where 40~K was used).
For those stars not in the FSC, upper limits are plotted based
on the catalog completeness
(0.2~Jy at 12, 25, 60~\micron\ and 1.0~Jy at 100~\micron, non-color corrected).
The stellar photosphere is shown as a Kurucz model for the appropriate
stellar type scaled to the 2MASS JHK fluxes.

The brightest source detected at submillimeter wavelengths,
HD~107146, is a G2 star with estimated age $\sim 100$~Myr.
It was strong enough to map and found to be marginally resolved
at 450~\micron. The maps and analysis of the SED were presented in 
Williams et al.\ (2004; hereafter Paper I).
Briefly, the fit to the SED implied a dust temperature of 
51K and a dust mass of $0.1 M_\oplus$. 
Based on the measured properties of the source, we predicted that 
the disk would be unusually bright in scattered light.  This was  
confirmed in recent imaging of the system with the Hubble Space Telescope 
(Ardila et al.\ 2004).  
The other two submillimeter detections in our survey are associated
with F8 stars: HD~104860 with an estimated age of $\sim 40$~Myr and
HD~8907 with an age $\sim 200$~Myr.

For these two systems, we fit the excess above the predicted photosphere
using a single temperature greybody with optical depth
$$\tau_\lambda=1-{\rm exp}[-(\lambda_0/\lambda)^\beta]$$ 
which has the desired asymptotic behavior, 
$\tau=1$ for $\lambda\ll\lambda_0$ and
$\tau=(\lambda_0/\lambda)^\beta$ for $\lambda\gg\lambda_0.$
The critical wavelength, $\lambda_0$, was set to 100~\micron\
for consistency with previous debris disk studies
(Dent et al. 2000; Wyatt, Dent \& Greaves 2003).
Further, because the photospheric excess was significantly detected
at only two wavelengths longward of 60~\micron\ in each case,  
we also fixed $\beta=1$ based on the typical SED slope for debris
disks (Dent et al. 2000) and protostellar disks (Beckwith et al. 2000).
The excess was then fit by varying the dust temperature,
$T_{dust},$ and scaling. The dust mass was determined from
$$M_{dust}={F_{850}\,d^2\over\kappa_{850}B_\nu(T_{dust})}$$
where $F_{850}$ is the SCUBA 850~\micron\ flux, $d$ is the
distance to the source, $\kappa_{850}$ is the dust mass absorption
coefficient, and $B_\nu$ is the Planck function.

The value of $\beta\simeq 1$ for disks is less than that observed
in the ISM ($\beta\simeq 2$) and
can be produced by a population of large grains
(e.g., Miyake \& Nakagawa 1993; Pollack et al. 1994).
Independent evidence for large grains in debris disks comes from
the lack of spectral features at mid-infrared wavelengths indicating
that the emission arises from grains $> 10~\micron$ in size (Jura et al. 2004)
and is consistent with a collisional origin for the dust
(e.g., Krivov et al.\ 2000).

Pollack et al. (1994) show that for spherical grains at 100~K,
$\beta\leq 1$ is only achieved for radii $\geq 1$~cm. The resulting values
of $\kappa_{850}$ range from 0.32 to 0.62~cm$^2$~g\e.
These are a factor of $3-5$ less than the canonical
value, $\kappa=1.7$~cm$^2$~g\e, that is commonly used in studies of
debris disks dating back to Zuckerman \& Becklin (1993).
On the other hand, porous, or fractal, dust grains can have
cross-sections that are more than an order of magnitude higher
than spheres per unit volume (Wright 1987).
Due to the uncertainty in the dust composition and structure,
we also adopt $\kappa=1.7$~cm$^2$~g\e\ for ease of comparison
with previous results but we note that the masses may be 
$3-5$ times higher than the values assumed here.

HD~104860 was undetected by IRAS at mid- to far-infrared wavelengths, 
but we measured emission in excess of the photosphere at both 850~\micron\
and 450~\micron\ indicating a cold, massive disk.
The SED is shown in Figure~1 where the photospheric excess is fit
by a range of $\beta=1$ greybodies 
with temperatures between 19 and 42~K.
The best fit has $T_{dust}=33$~K and $M_{dust}=0.16~M_\oplus$.
Table~2 summarizes the corresponding range of dust masses and excess 
luminosities.
The lack of an IRAS detection and the low temperature might suggest
that the submillimeter flux is due to 
Galactic cirrus emission but the high latitude of the source, $b=50^\circ$,
makes this unlikely.
The low temperature of the dust surrounding HD104860 is similar to 
that of three other cold disks that have been reported
around companions to early type stars (Wyatt et al.\ 2003).

HD~8907 was detected by IRAS at both 12~\micron\ and 60~\micron\ 
and by ISO at 60~\micron\ and 87~\micron\ (Silverstone 2000).
The $12\micron$ flux is photospheric, whereas the longer wavelength 
fluxes represent a strong excess.
In the submillimeter, an excess is detected at 850~\micron\
and marginally ($2\sigma$) at 450~\micron.
The $\beta=1$ greybody fit to the excess above the photosphere
is tightly constrained by these points with a
temperature $T_{dust}=48$~K,
%luminosity $L_{dust}=5.4\times 10^{-4}~L_\odot$,
and mass $M_{dust}=0.036~M_\oplus$ (Figure~2).

The SEDs of the remaining ten non-detections are shown in Figure~3.
Three objects, HD~17925, HD~35850, and possibly HD~166435
show mid-infrared excesses above the stellar photospheres, but
lack detections longward of 100~\micron.
The 3-$\sigma$ upper limits to the 850~\micron\ flux and
inferred disk masses are presented in Table~1.
The flux upper limits were converted to mass upper limits using 
an average temperature for debris disks around solar type stars.
Since stars with luminosities $L_* < 10~L_\odot$ have dust 
temperatures within a narrow range 33 to 56~K, averaging 50~K 
(Wyatt et al.\ 2003; Sheret et al.\ 2003; Sylvester et al.\ 2001; 
Liu et al.\ 2004; Greaves et al.\ 2004b), we 
therefore adopted an average temperature of 
$\langle T_{dust}\rangle = 50$~K in estimating the mass upper limts. 
Note that all the upper limits are less than or equal to the most
massive disk in our sample, HD~104860,
and five are less than the least massive disk, HD~8907,
and therefore these non-detections provide statistically useful information.
The range in source distances of course enters into the mass 
upper limits, but the sensitivity of the observations is nevertheless 
adequate to show that there is a range of disk masses present around 
stars of similar age.
For example, for HD17925 at a distance of 10 pc and an age of 100 Myr,
our mass upper limit is $0.005 M_\oplus$.
In comparison, the more distant source HD~107146 (at 28.5pc)
has a similar age and a mass that is 20 times
larger.  These results provide further evidence for intrinsic diversity
in the evolution of the circumstellar dust content of solar-type stars 
(cf. Meyer et al. 2004).

\subsection{CO (3--2)}
We can convert the upper limits on the CO(3--2) integrated intensities 
for HD107146 and HD104860 to upper limits on the CO gas mass 
assuming LTE and an excitation temperature for the gas  
(e.g. Scoville et al.\ 1986). 
For HD107146, we assume an excitation temperature of $T_{\rm ex}=50$~K  
based on the estimated dust temperature for the system (Paper I) 
and a FWHM for the line of $1.3-2.3$~\kms, which is appropriate for gas 
at $\sim 30-100$ AU in orbit around a $\sim 1 M_\odot$ star 
at an inclination of 25$^\circ$ (Ardila et al.\ 2004). 
For the more conservative limit corresponding to a FWHM of 2.3~\kms, 
the $3\sigma$ upper limit to the mass of gaseous CO is 
$M_{\rm CO}<2\times 10^{-6}~M_\oplus$. 
For HD104860, we assume $T_{\rm ex}=30$~K  
based on the estimated dust temperature for the system (Table 2). 
Since the low dust temperature is consistent with dust emission 
from large disk radii ($\gtrsim 100$ AU), 2~\kms\ FWHM is 
a reasonable line width. 
With these assumptions, the $3\sigma$ upper limit to the mass of 
gaseous CO is $M_{\rm CO}<1.3\times 10^{-5}~M_\oplus$. 
These limits vary by less than 50\% for 
excitation temperatures in the range $T_{\rm ex}=25-100$~K.

A critical issue is the conversion from CO to total gas mass.  Processes
that can reduce the abundance of gas phase CO include condensation on
grains and photodissociation 
(e.g., Dent et al.\ 1995; Kamp \& Bertoldi 2000).  
If the upper limit on the CO mass is to be consistent with a 
primordial gas-to-dust ratio of 100 
(i.e., a total gas mass of $10\Mearth$), 
these processes must be extremely efficient given the large dust mass
and our upper limit on the CO mass.   
For example, for the dust mass of $0.10\Mearth$ measured for HD107146 
(Paper I; see \S3.1), 
a gas-to-dust ratio of 100 implies a CO to total hydrogen abundance
of $n_{\rm CO}/n_{\rm H_2}=2\times 10^{-8}$ (assuming a 10\% fraction 
of Helium by number), 
several orders of magnitude below the interstellar value of
$n_{\rm H_2}/n_{\rm CO} \simeq 10^4$ (e.g., van Dishoeck et al.\ 1993).

For the condensation of CO on grains, an important issue is the 
substrate onto which CO is adsorbed. 
The binding energy of CO onto water ice is 1740 K, whereas the
CO--CO surface binding energy is significantly lower, 960 K
(Sandford \& Allamandola 1988, 1990).
In their study of well-known debris disks, 
Kamp \& Bertoldi (2000) assumed CO adsorption onto a water ice 
substrate, in which case where the critical grain temperature for 
condensation turns out to be $\sim 50$~K. 
In contrast, studies of CO condensation in protoplanetary disks 
have typically assumed a CO substrate  where the grain condensation 
temperature is $\sim 20$~K (e.g., Aikawa et al.\ 1996). 
The relevant substrate likely depends on the thermal history of 
the grains in the disk.  

For example, if the grains are imagined to be heated and 
rapidly cooled, gas phase water and CO will condense together, 
with CO binding to water ice. 
In contrast, if the grains are 
heated and slowly cooled, water will condense first, 
followed by CO.  In the latter case, the abundance of CO in the gas 
phase relative to the number of grains is sufficient to establish 
a CO substrate.  That is, for a gas-rich disk in which the 
gas-to-dust mass ratio is 100, with all of the dust in grains of 
radius $a,$ and $n_{\rm CO}/n_{\rm H}\sim 10^{-4}$, there are 
$\sim 10^{13}$ CO molecules per $10\micron$ dust grain.
In comparison,
there are $4\pi a^2 n_s$ binding sites available on the grain,
where $n_s\simeq 1.5\times 10^{15}\persqcm$
(Tielens \& Allamandola 1987),
or $\sim 2\times 10^{10}$ sites per monolayer.
Thus, the condensation of only a very small fraction ($< 1$\%)
of the CO is needed to establish a CO substrate.
In this case, because of the low CO--CO surface binding energy, 
the critical temperature for condensation will be lower than 
that estimated by Kamp \& Bertoldi (2000). 

For HD107146, the measured temperature of the emitting grains is high 
($\sim 50$~K), comparable to the evaporation temperature for even 
CO on water ice.  If the emitting grains are large and therefore 
at the blackbody temperature, condensation of CO on grains is 
not significant.  
Such grains would be located close to the star, at $\sim 30$ AU. 
However, if the $\sim 50$~K grain temperature characterizes 
the emission from a population of small (non-blackbody) grains  
located farther from the star,  
a larger, cooler population of grains, if also present,
could potentially deplete CO from the gas phase.
A distance of 100 AU is a reasonable maximum distance for the 
emitting grains given the spatially resolved continuum emission at 
450\micron\ (Paper I).  At this distance from the star, grains 
$\gtrsim 100\micron$ will be at the blackbody temperature of 28 K.  
Significant condensation will begin to occur when the adsorption rate
onto grains exceeds the evaporation rate.
For condensation onto a CO substrate, evaporation
will keep CO from condensing further up to a gas phase CO density of 
$n_{\rm CO}< 3\times 10^{9}\percc.$
This density limit exceeds not only the density expected
at the midplane of the minimum mass solar nebula,
but also the much lower densities expected in the lower mass
HD107146 disk.
Thus,  
a gas-rich disk is not expected to experience significant CO 
condensation for grains located anywhere from the inner radius 
of $\sim 30$~AU out to $\sim 100$~AU. 

For HD104860, the temperature of the emitting grains is 
lower, $\sim 30$~K, and the uncertainty in temperature allows 
the possibility of a temperature as low as 19~K (Table 2). 
Given the possibility that the emission arises from a 
population of relatively small (non-blackbody) grains and that 
larger, cooler grains may be present, it is difficult to rule 
out the possibility of significant CO condensation in this case. 

Another process that can reduce the gas phase CO abundance
is photodissociation.  Kamp \& Bertoldi (2000) have shown that 
FUV irradiation by the star and the interstellar radiation field can
produce significant underabundances of CO in the disks around
Vega and $\beta$ Pic.  Using these results, they argue that the
non-detection of CO in these systems may be consistent with a
primordial gas-to-dust ratio in these systems.
Other studies of debris disks surrounding both A-type and 
solar-type stars have previously argued that CO has been significantly 
dissociated and that the upper limits on CO mass are consistent 
with primordial gas-to-dust ratios 
(Dent et al.\ 1995; Greaves et al.\ 2000). 
A later spectral-type star such as HD107146 is believed to be 
a negligible source of dissociating flux compared to the interstellar
radiation field (Kamp \& Bertoldi 2000; Greaves et al.\ 2000).

Following Dent et al.\ (1995) and Greaves et al.\ (2000), we can
use the photodissociation models of van Dishoeck \& Black (1988)
to estimate the extent to which CO will be dissociated in
the HD107146 disk.  
These models show that the relative abundance of CO is a function 
of both the density within and the column density though the medium. 
For a gas-to-dust ratio of 100, the expected 
disk gas mass is $10 \Mearth$. 
This corresponds to a vertical hydrogen column density
of $\sim 4\times 10^{21} - 4\times 10^{22}\persqcm$ for gas 
uniformly distributed in a disk 30--100 AU in radius, and a mean 
density of $n_{\rm H}\sim 10^7-10^9\percc$ assuming a thermal 
scale height.  At such large column densities,
CO molecules are expected to be well-shielded.  The models of
van Dishoeck \& Black (1988) predict that at these column densities 
the ratio of the CO and hydrogen column densities is
$\gtrsim 10^{-4},$ several orders of magnitude larger than the ratio of
$2\times 10^{-8}$ required for a primordial gas-to-dust ratio.

Thus, our CO mass limit appears to completely rule out a
primordial gas-to-dust ratio for HD107146.
How low might the actual gas content be?
Once the total hydrogen column density of the disk begins to drop below
$10^{21}\persqcm$, the CO molecules are exposed to a higher
photodissociating flux and the CO abundance relative to hydrogen 
declines below the well-shielded value.
For a gas-to-dust ratio of order $\sim 1$, we would be in the 
regime where the expected hydrogen column density and hydrogen 
number density are $N_{\rm H}\sim 10^{20}\persqcm$ 
and $n_{\rm H}\sim 10^5-10^7\percc$, respectively.  
Model T6C of van Dishoeck \& Black (1988; $n_{\rm H}\sim 10^4\percc$) 
provides the closest match to the expected range of density. 
At hydrogen column densities $N_{\rm H} \sim 10^{20}\persqcm,$ 
the model predicts that the ratio of CO and H$_2$ column 
densities is $N_{\rm CO}/N_{\rm H_2} \sim 10^{-5}$ or a mass ratio of 
$M_{\rm CO}/M_{\rm H_2} \sim 10^{-4}.$  Thus the model implies 
that our upper limit on the CO mass of $2\times 10^{-6}M_\oplus$ 
corresponds to an upper limit on the total gas mass of 
$\sim 0.02M_\oplus$, or a gas-to-dust ratio of $< 1$,  
significantly below the primordial value.

One of the possible flaws in this argument is that the 
van Dishoeck \& Black models used to interpret the results 
were developed to address the 
conditions in interstellar clouds rather than those in 
circumstellar disks.  So, for example, they describe lower 
density conditions than may apply to HD107146.  They also 
do not include the possible effects of additional processes, 
such as advanced grain growth and stellar X-ray and UV excess 
irradiation, on the chemical abundance 
structure of the medium.  
We have also assumed that the gas and dust temperatures are 
equal, an assumption that should be examined theoretically. 
An improved analysis would be possible 
with the recent models of Gorti \& Hollenbach (2004) which 
include such processes and more directly address the thermal 
and chemical structure of residual gas disks.  This is a topic 
for future investigation.

\section{Discussion}
 
\subsection{Dust Disk Radii and Masses}

Single temperature greybody fits like those employed here are commonly 
used to interpret sparsely sampled SEDs of debris disk sources. 
Taken at face value, such a fit implies
that the emission arises from a limited range of disk radii, in particular
that the dust distribution has an inner hole.  
An inner hole in the dust distribution could arise in several different ways.
For example, in a disk composed of icy grains, grain destruction at 
the water ice evaporation radius can produce an inner hole in the 
dust distribution.  
While this situation may apply to disks surrounding 
luminous (A- and B-type) stars that have dust temperatures $\sim 100$ K, 
similar to the water ice sublimation temperature 
(e.g., HR4796A---Jura et al.\ 1998), 
the low dust temperatures of the sources 
detected in our survey (30--50 K) suggest that this situation is not 
relevant here.  

Primordial gas-rich disks may also develop optically thin regions as a 
result of grain growth and planetesimal formation.  Since collisional 
evolution is more rapid at small radii, optically thin inner holes 
are expected to develop.  In the gaseous outer disk, large dust 
masses can be maintained since gas drag can prevent grains from 
developing eccentric orbits and substantial relative velocities, 
and destructive collisions are thereby minimized. 
Such gas-rich outer disks with optically thin inner holes have been proposed 
to explain the large excesses detected in young disks systems such as 
TW Hya ($\sim 10$ Myr; Calvet et al. 2002) and CoKu Tau/4 
($\sim 1$ Myr; D'Alessio et al. 2004). 
Dusty disks that possess a residual gas component 
($M_{\rm gas} \lesssim 10 M_{\rm dust}$) 
can also develop ring-like structures 
as a result of  gas-grain coupling in the presence of 
stellar radiation pressure.  This situation has been proposed to explain 
the ring-like structures seen in 
systems such as HR4796A and HD141569 (e.g., Takeuchi \& Artymowicz 2001). 

The theoretical expectation that gaseous outer disks 
photoevaporate on short 
timescales ($\sim 10$ Myr; Hollenbach, Yorke, \& Johnstone 2000) 
and the lack of observations to the contrary both suggest that 
the above situations are unlikely to apply to 
our sample (age = 10--200~Myr), 
although they are difficult to rule out. 
For solar mass stars, 
irradiation by the central star is thought to be capable of 
photoevaporating away the gaseous component of the outer region ($r > 10$ AU) 
of a minimum mass solar nebula on a timescale of 10 Myr, 
if the Lyman continuum flux from the star 
remains at classical T Tauri levels for this length of time 
(Hollenbach et al.\  2000).  
Since the Lyman continuum flux level is unlikely to continue at such 
levels for longer than 10 Myr, 
it may be difficult to dissipate more massive disks 
through photoevaporation by the central star alone.  
TW Hya may be one such example of a $\sim 10$ Myr old star with a substantial 
outer gas disk that has survived an earlier phase of energetic 
photoevaporation (Kastner et al.\ 1997; Qi et al.\ 2004 and references 
therein). 
Other processes such as viscous dissipation may then dominate the 
dissipation of the disk. 

Observational support for the rapid dissipation of gaseous outer disks 
comes from the results of searches for gas in 1--10 Myr sources 
(e.g., Zuckerman, Kastner, \& Forveille 1995) and known debris disk systems  
(e.g., Greaves et al. 2000)
The limited data available for our sample (sec.\ 3.2) indicate 
an upper limit to the gas mass for HD107146 that is less than 
the dust mass in the system, but there is no significant constraint 
on the gas content of the rest of the sample.  Even for HD107146, 
although our results apparently rule out a primordial gas-to-dust 
ratio for the system, the upper limit on the gas mass may still allow 
gas-grain coupling and radial migration of grains to play a role 
in sculpting the dust disk.

The possibility that these studies, which primarily 
do not trace H$_2$ directly, have underestimated the disk gas content 
has been raised by 
the surprising detection with ISO of abundant ($\sim 1 M_J$) 
H$_2$ in $\sim 20$ Myr old debris disk systems (Thi et al.\ 2001).  
Spitzer offers an opportunity to confirm these measurements and 
to make similar measurements for a wider range of debris disk sources. 
Some uncertainty already surrounds the ISO results given since 
ground-based searches for mid-IR $H_2$ emission from younger 
T Tauri stars do not confirm the ISO detections
(Richter et al.\ 2002; Sheret et al.\ 2003; Sako et al.\ 2005). 
In addition, 
recent Spitzer observations do not confirm the reported H$_2$ line 
fluxes for $\beta$ Pic (C. Chen, personal communication). 
Whether the implied lower gas masses are representative of the general 
situation for debris disk systems at 20 Myr and what constraints are 
placed by the observations on the disk gas mass 
are interesting issues for the future. 

Given the existing uncertainties regarding the gas dissipation 
timescale in disks, it is difficult to rule out the possibility that 
sources in our sample may have significant gaseous outer disks; 
gas-grain interactions may therefore contribute to the structure that 
is deduced from the observed SED.  
For the rest of the paper we assume, consistent with convention, 
that at the relatively advanced age of the sources in our sample (10--200 Myr) 
little residual gas remains in the outer 
disk region.  
Sensitive future searches for gas in these or similar systems, 
such as those to be made with Spitzer as part of FEPS, 
may spark a reexamination of this issue. 

In the absence of a substantial gaseous component to the disk, 
grain lifetimes are affected by both Poynting-Robertson (PR) drag 
and particle collisions.  Collisions are expected to dominate the 
destruction of grains at high disk masses 
(e.g., Dominik \& Decin 2003; Wyatt 2005).  As described 
by Backman \& Paresce (1993), the grain mutual collision time is 
$t_{\rm coll} \sim P/8\sigma$ where $P$ is the orbital period and 
$\sigma$ is the fractional 
surface density (the surface filling factor) in grains.  
This estimate assumes that grains are on orbits that are sufficiently 
inclined that the grain encounters the full surface density of the 
disk approximately twice per orbit as it oscillates through the disk 
plane.  The estimate also includes tangential collisions as 
destructive events, i.e., the collisional crosssection is 4 times 
the crosssectional area of individual grains. 
For a total dust mass $M_{dust}$, spread out over an area $\pi R_{dust}^2$, 
internal grain density $\rho_i$, and average grain radius $a$,  
$$t_{\rm coll} = 3000\,{\rm yr}
\left(R_{dust}/30 {\rm AU}\right)^{7/2}
\left(M_{dust}/0.1 M_\oplus\right)^{-1}
\left(a/1 {\rm mm}\right).$$
Thus, for the sources in our sample with detected dust disks 
($M_{dust} \sim 0.1 M_\oplus$; Table 1), 
the collision timescale is significantly shorter than the age of 
the systems, and the observed grains must be replenished by collisions 
between larger bodies. 

In comparison, the PR drag time for blackbody grains is 
$$t_{\rm PR} = 640\,{\rm Myr} 
\left(a/1 {\rm mm}\right)   
\left(R_{dust}/30 AU\right)^2    
\left(L_*/L_\odot\right)$$    
where $L_*$ is the stellar luminosity.
For the millimeter-sized grains deduced from the submillimeter 
observations, collisions therefore dominate as the destruction process 
down to dust masses  $M_{dust} < 10^{-6} M_\oplus$, well below the 
dust masses inferred for the systems studied here. 
At dust masses above $\sim 10^{-6} M_\oplus$, 
grains are destroyed by collisions before 
PR drag can cause them to migrate very far from where where they 
were created.  As a result, the region of the disk that contains 
the collisional debris is also the region of the disk in which the 
colliding planetesimals that produce the debris are located. 

We can estimate the orbital radii of the collisional debris for the 
sources detected in our sample assuming that the grains responsible 
for the emission are large and absorb and emit as blackbodies.  The 
$\sim 50$ K dust temperature of HD107146 and HD8907
translates into grain orbital distances of approximately 
30 and 48 AU respectively.
The lower ($\sim 30$ K) temperature of the dust in the HD104860 
system locates the collisional debris at $\sim 110$ AU.  We therefore 
infer the presence of a population of colliding planetesimals at these 
radii. 
The presence of a large planetesimal population in these young 
solar analogue systems at orbital radii similar to that of the Kuiper 
Belt in our solar system (e.g., Jewitt \& Luu 2000, PPIV) suggests 
their utility in understanding the early evolution of Kuiper Belt 
systems.

\subsection{Comparison With Models}
Several recent models for the collisional evolution of planetesimal 
disks suggest possible interpretations for these data.  
Dominik \& Decin (2003) describe the evolution of a ring of planetesimals 
located at a fixed distance from the star. 
Assuming that the ring is in equilibrium, in the sense that 
grain production and destruction rates balance, the mass in small 
grains evolves as $M_{dust} \propto t^{-1}$ if grain destruction is 
dominated by mutual collisions and $M_{dust} \propto t^{-2}$ if 
grain destruction is governed by PR drag
(see also Spangler et al.\ 2001).
The former case would apply at early times when the disk mass is 
large and collisions are frequent; the latter would dominate at 
late times when collisions become infrequent enough the PR drag 
begins to dominate. 
The model is flexible in that it does not specify how the 
eccentricity and inclination distributions of the planetesimals are 
determined, i.e., how and when the population is stirred. 

In a complementary, more prescriptive model, Kenyon \& Bromley (2004) 
follow the evolution 
of a disk of planetesimals undergoing planet formation.
They find that once embedded objects grow to $>1000$ km in size, they 
stir up the surrounding planetesimals, generating a collisional cascade 
and accompanying debris.  
That is, the planetesimal disk is ``self-stirred'' by 
embedded protoplanets. 
Since the evolution of the disk is dominated by collisions, 
the growth timescale scales as the collision time 
$t_{coll} \propto P/\Sigma$ where $P$ is the 
orbital period and $\Sigma$ is the disk surface density in planetesimals.  
Since both the orbital speed and disk surface density decline with 
disk radius, the growth time is an increasing function of disk radius.
Thus, a wave of planet formation propagates outward in the disk: as time
progresses, the formation of embedded protoplanets excites collisions
among the less massive bodies, thereby generating dust debris at
successively larger characteristic radii.

In this model, the total mass in small grains in the disk remains 
remarkably constant with time (for perhaps as long as 1 Gyr),
and subsequently decays as $\propto t^{-n}$ where $n=1-2$.
In contrast, the reprocessed luminosity emitted by the collisional 
debris begins to decline at a much earlier time (after only 1--10 Myr). 
This is because as the wave of planet
formation moves outward, grains of a given size subtend increasingly
smaller solid angles as they are located at larger distances from the star.
Although the models calculated by Kenyon \& Bromley are for more massive 
stars ($M_*=3 M_\odot$) than those in our sample, we can use the  
scaling of $t_{coll} \propto P/\Sigma$ to estimate the behavior 
that would be obtained for lower mass systems.  

Comparing our results with these models, we find that 
the deduced size of the dust disks for the detected systems and their  
nominal stellar ages are roughly consistent with predictions for  
the evolution of dust production in collisional planetestimal 
disks.  The Kenyon \& Bromley (2004) model predicts that in a disk with 
a mass comparable to that of the minimum mass solar nebula that 
surrounds a solar mass star,   
it takes $\sim 300$ Myr for planet formation to reach 40 AU and to 
produce a local maximum in the dust production rate at that distance. 
This is comparable to the sizes (30--50 AU for HD8907 and HD107146) 
and ages (200 and 100 Myr respectively) of the sources studied here.

The derived dust masses for the sources detected here are
$0.03-0.16M_\oplus$.  These dust masses are 2--8 times the mass 
in small grains that is predicted by Kenyon \& Bromley (2004) to 
reside in a disk with the 
mass of the minimum mass solar nebula that is undergoing planet 
formation ($0.01-0.02 M_\oplus$).
In the Kenyon \& Bromley (2004) study, the mass in collisional debris was
found to be remarkably independent of the bulk properties of the
planetesimals and initial conditions other than disk mass.  
Instead, the dust mass was found to depend primarily on 
(and to be proportional to) the initial mass
of solids in the disk.  Interpreted in this context, the larger
disk masses measured here suggest that they arise from more massive
versions of the minimum mass solar nebula.

The Kenyon \& Bromley (2004) calculation 
provides a way in which to deduce the mass of planetesimals
that currently resides in the disk 
(see also Dominik \& Decin 2003).  
Their results predict that at an age of 100 Myr
$\sim 4\times 10^{-4}$ of the
mass in solids is in the form of grains 1 mm and smaller.  
Thus the measured mass in small grains implies a current mass 
in parent bodies of
$\sim 85 M_\oplus$ for HD8907,
$\sim 250 M_\oplus$ for HD107146, and
$\sim 400 M_\oplus$ for HD104860. 
They also find that over the age range of the sources 
in our sample, the mass in small dust grains is roughly half that 
of the initial mass of solids in the disk.  
Thus the initial mass in solids in the outer planetesimal disks 
of the detected sources would be twice as large 
($\sim 170- 800 M_\oplus$). 
For comparison, 
the mass of the Kuiper Belt in our solar system today is
a mere $\sim 0.1 M_\oplus$; 
in primordial times, 
the minimum mass solar nebula in the entire radial range
30--150 AU contained $\sim 100 M_\oplus$ and
several 10s $M_\oplus$ are thought to have been needed in the 
30-50 AU range in order to form Pluto and QB$_1$-type 100m KBOs
(Stern \& Colwell 1997; Kenyon \& Luu 1998).

Since our sample focuses on solar mass stars, it is interesting to
ask what relevance these stars might have to our solar system.
The large dust masses measured for the submillimeter sources
detected in our study indicate initial disk planetesimal masses
2--10 times that of the minimum mass solar nebula.  Such
massive disks favor the production of giant planets like those in the
solar system.  Massive disks make it easier to form Jupiter with
the possibly short lifetime of the gaseous disk (Thommes et al.\ 2003).  
So these systems may be reasonable analogues
of the young solar system.

The inferred masses of the planetesimal disks in the current epoch 
are massive enough to play a
role in determining the architectures of planetary systems.
Planetesimal disks $\sim 100 M_\oplus$ are expected to induce
giant planets to migrate significantly on $\sim 100$ Myr timescales
(Hahn \& Malhotra 1999),
since the clearing of a residual planetesimal disk by planets involves
a significant exchange of orbital energy and angular momentum.
In the case of our solar system, the interaction between a
$10-100 M_\oplus$
residual planetesimal disk and the giant planets is believed to have
caused Jupiter to migrate inward and Saturn, Uranus and Neptune to
migrate outward (e.g., Hahn \& Malhotra 1999).
In particular, the significant outward migration of Neptune
(by $\sim 7$ AU) is thought to have sculpted the Kuiper Belt and
pushed it out to the orbital distance range that it currently occupies
(40--50 AU; Hahn \& Malhotra 1999; Levison \& Morbidelli 2003).
The disks detected here appear to be sufficiently massive to induce 
similar migration among any giant planets that may be present. 

When interpreted (literally) in the context of the Kenyon \& Bromley
(2004) model, the collisional debris that we have observed
surrounding the $\sim 100$ Myr sources in our sample
suggests that we are witnessing the formation of
$\sim 1000$ km bodies at distances of 30--70 AU around nearby stars.
For HD107146 and HD8907, which have debris disks at $\sim 30-50$ AU,
the large bodies that are forming may correspond to
proto-Plutos.
These disks are comparable to the size of the 
Kuiper Belt in the solar system, which appears to have an edge at 
$\sim 50$ AU (Allen, Bernstein, \& Malhotra 2001; Chiang \& Brown 1999). 
The possibility of a primordial Kuiper Belt extending beyond 
50 AU has been discussed in the literature 
(e.g., Stern 1995), and indeed  
HD104860 is one example of a young (40 Myr) solar-mass star with 
a planetesimal disk that appears to extend much beyond 
known extent of the Kuiper Belt in our solar system 
(to $R_{dust} \sim 110$ AU) and is massive enough at that distance 
to enable the formation of a massive perturbing object.

\subsection{The properties of disks detected at 850~\micron}

Since the inferred properties of the debris disks (planetesimal 
and initial disk masses) and potential implications discussed above 
rely on the calculations of Kenyon \& Bromley (2004), one 
might wonder about the applicability of these models to the 
sources that we observed. 
For example, a possible limitation of the Kenyon \& Bromley study 
is that it does not include the possible formation of giant planets 
at small radii and their impact on the collisional evolutionary 
history of an outer planetesimal disk.  
The presence of shepherding giant planets has often been invoked 
to explain the ring-like structures, observed or inferred, in 
disk systems over a range of ages, from $\sim 1$ Myr (e.g., 
CoKu Tau 4; D'Alessio et al.\ 2004) to $\gg 100$ Myr (e.g., 
$\epsilon$ Eri; Greaves et al.\ 1998). 
In particular the likelihood of a diversity of giant planet 
architectures, similar to that seen in the extrasolar planet 
population (Marcy \& Butler 1998), may lead to a diversity of 
collisional evolutionary histories that depart from those predicted 
by self-stirred models. 
One way to explore the extent of such resultant diversity and 
the applicability of the models to observations of outer 
planetesimal disks is to compare the model predictions with 
the collective properties of debris disks detected at submillimeter 
wavelengths. 

We therefore collated the submillimeter excess sources from the 
literature and combined them with the results of our 
survey. 
In practice, almost all detections of debris disks at sub-millimeter
wavelengths have been made with the SCUBA camera on the JCMT
at 850~\micron  
(Greaves et al. 2004a, 2004b; Wyatt et al. 2003, 2004; 
Sheret et al. 2004; Liu et al. 2004). 
We henceforth refer to this sample as the 850~\micron\ disks.
In the following sections we examine compare the 
masses, radii, and ages for the sample with the predictions of 
recent models for the collisional evolution of planetesimal 
disks.

\subsubsection{Disk Mass Evolution}
Figure 4 plots the dust masses and ages of submillimeter excess 
sources from the literature 
combined with the detections 
reported here.  For these sources, 
the SCUBA fluxes were converted to masses assuming $\kappa=1.7$~cm$^2$~g\e\ 
and dust temperatures from the literature.  Non-detections are 
indicated in Figure 4 as 3-$\sigma$ upper limits; these 
are calculated assuming a dust temperature of 50 K.  
The sources shown cover a range of spectral types, from B7 to M4.  
The resulting plot 
is similar to those that have appeared previously in the
literature (e.g., Wyatt et al.\ 2003; Liu et al.\ 2004).
Our detections follow the trend established by the earlier 
measurements, which shows a decline in the dust mass as a function 
of age.  Such a decline is expected to result from the collisional
erosion of a planetesimal disk.

Several significant selection effects complicate the interpretation 
of this plot. For example, the detected sources are mostly early type 
stars, which is likely a result of sensitivity considerations. 
If these more massive stars form with more massive disks, 
they will have preferentially larger dust masses at any given age, 
making them easier to detect.
The higher luminosity of the central star also enhances 
the detectability of any surrounding dust. 
Another selection effect is that 
there are few sources $\le 10$ Myr in age 
that are near enough to be readily detectable at low 
masses.  (The sources in the TW Hya and $\beta$ Pic moving groups are 
notable exceptions.)  
Since such sources are typically located at larger distances than the older 
sources that have been detected, the lack of sources in the lower 
left region of the plot may simply result from sensitivity limitations. 
Thus, although there are no sources with low measured dust masses at 10
Myr age, there are many upper limits.  

The upper right region of the plot is also not well populated,  
a property that seems difficult to attribute to a selection effect.  
In addition, some of
measurements made here (in the 10--100 Myr range) fall below the
mean mass--age relation. 
This suggests that the detected sources
represent the upper envelope of a broader distribution.
This is consistent with the interpretation discussed earlier that the
disks detected here arise from 
more massive versions of the minimum mass solar nebula.  
%disks a few to 10 times the mass of the minimum mass solar nebula.  
Much more massive disks are not expected since 
they would be gravitationally unstable. 

To investigate the constraints that these data place on the evolution of
debris disks, in particular the information contained in the many 
non-detections, we carried out a
survival analysis (Feigelson \& Nelson 1985) using the ASURV Rev 1.2
package (LaValley, Isobe, \& Feigelson 1992). We divided the sample into a
`young' and `old' group based on a dividing age, $t_0$, and then computed
the probability that they were drawn from the same population.
The probabilities of similarity decreased from $7-12\%$
at $t_0=100$~Myr to $2-3\%$ at 200~Myr and down to $1-2\%$\footnotemark
\footnotetext{The precise probabilities depended on the particular
statistical test used and the weighting that they give to the upper limits.}
at 500~Myr.
Thus, there appears to be a significant difference between the `young' 
and `old' groups at a dividing age of $t_0\sim 200$ Myr.  That is, the 
mass in small grains decreases significantly on a $\sim 200$ Myr timescale.

This result is complementary to that obtained by Carpenter et al.\ (2004). 
Using survival analysis, they compared the submillimeter excess detections 
and upper limits obtained for a younger sample of FEPS sources with
the submillimeter properties of sources in Taurus.  They found 
no statistically significant difference between the Taurus disk population 
and 3--10~Myr systems,
somewhat weaker evidence for a difference with the 10--30~Myr population, 
and a clear difference between the Taurus disk population and 
30--100~Myr systems. 
The results presented here suggest the decline to lower dust masses 
continues at ages beyond $\sim 200$ Myr. 

Although the small size (and heterogeneous nature) of the submillimeter 
sample makes it difficult to constrain the evolutionary behavior of 
the sample in greater detail, 
the dashed line in Figure 5 shows that the measurements can be fit 
by an overall 1/(Age) dependence of dust mass vs.\ age with a 
significant spread about this trend. 
This overall trend is the behavior expected for a ring 
of planetesimals in which grain destruction is governed by collisions 
(Dominik \& Decin 2003). 
This slope is also consistent with that predicted in the 
Kenyon \& Bromley (2004) models for ages $>10$ Myr. 
Some of the spread in the observed dust mass at a given age may be 
due to the range of stellar spectral types present in the sample. 
For example, 
Dominik \& Decin (2003) show that a range of spectral types 
A0V to K3V introduces a spread of about an order of magnitude 
in fractional dust luminosity at an age of 1 Gyr, 
a range comparable to the spread found here. 
Additional heterogeneity in the sample, such as a range of 
initial disk masses or a diversity of giant planet formation 
histories, would also contribute to 
a range in disk mass at any given age. 

A caveat to these results is the possibly significant uncertainty
in the estimated ages for the stars in the sample.  The ages in
Table 1, which derive from those estimated by FEPS, are typically
uncertain by factors of 2--3.  Similar errors are likely to characterize
the ages of the sources from the literature.  Note, however, that 
larger age uncertainties may characterize some sources.  For example,
subsequent to their selection for this study, HD88638 and HD166435
were both dropped from the FEPS sample because their Li strengths
indicated ages older than the $\sim 10$ Myr age estimated initially
for these sources.  So while typical age errors of a factor of 2--3
would not significantly affect our conclusions, this issue may need to
reexamined if larger errors are eventually found to be common.

Since the survival analysis formalism used above does not account for 
errors in the variables, we tested the sensitivity of the inferred decline 
in mass with age by varying individual stellar ages by up to a factor 
of 3.
We found that only when several systems crossed over the 200~Myr
timeline did the statistics change substantially. The effect was
to broaden the dividing age $t_0$ into an age range of $70-300$~Myr 
over which an ``old'' population is significantly less massive than a 
``young'' population.  The two populations and the approximate
border between them are shown in the two-tone greyscale in Figure~4.

\subsubsection{Disk Temperatures and Inner Radii} 
The Kenyon \& Bromley (2004) models make the simple prediction that
the characteristic dust radius increases with age in a given 
system if self-stirring dominates 
the collisional evolution of planetesimal disks. 
We therefore compared the stellar ages with the dust radii inferred 
for the sample.  
The dust temperatures of the sources, either as measured in section 3 
or from the literature, were converted to characteristic dust radii 
assuming blackbody grains:
$R_{dust,bb}=7.8\times 10^4\,{\rm AU}\,(L/L_\odot)^{0.5}\,T_{dust}^{-2}$.
Stellar ages were taken from the literature. 

Figure 5 plots the derived disk radius as a function 
of stellar age.  
No obvious correlation is found for the sample as a whole.  
Since the large range in stellar spectral types may obscure a 
trend, it would be preferable to compare the properties of a 
sample spanning a smaller spectral type range.  Among the A stars 
in the sample no strong trend is found although sample  
statistics are admittedly poor. 
We might attempt to include all of the sample in a comparison by 
making an assumption about how the dust evolution depends on the 
stellar mass.  If the time for debris production to reach a 
given radius $R_d$ is related to the collision timescale 
$t_{coll} \propto P/ \Sigma$ where the orbital period 
$P\propto R_d^{3/2}M_*^{-1/2}$ 
and the surface density in planetesimals follows the slope 
inferred for the solar nebula  
$\Sigma \propto R_d^{-3/2},$ 
then 
$t_{coll} \propto R_d^3 M_*^{-1/2}.$
Assuming $t_{age} \sim t_{coll}$, we would expect 
$t_{age} \propto R_d^3 M_*^{-1/2}$. 
Since the stellar mass of the sample varies by only 6 in the 
sample, scaling by $M_*^{1/2}$ produces only a minor variation 
on Figure 5. 
Another hypothesis might be that the initial planetesimal disk 
mass scales with stellar mass so that 
$\Sigma \propto M_* R_d^{-3/2}$. 
In this case 
$t_{coll} \sim R_d^3 M_*^{-3/2}.$
The stronger dependence on $M_*$ tightens up the distribution 
slightly but the spread is still large and no strong trend is 
apparent.

There are several caveats to these results. 
For one, we have assumed that the grains responsible for submillimeter 
excesses are large enough that they emit as blackbodies.  This assumption 
is supported by recent results from Spitzer which show that 
debris disks typically do not show emission features, suggesting that 
the grains responsible for the excess are large $> 10\micron$ 
(Jura et al.\ 2004).  Similar 
spectroscopic results for the specific objects studied here would 
put our assumption on firmer footing. 

Perhaps more significantly, sparsely sampled SEDs do not allow 
detailed fits that independently constrain the dust temperature and 
$\beta$.  (The spectral slope $\beta$ is a function of grain composition 
and shape as well as the grain size distribution.) 
As a result, we have typically assumed $\beta=1$ in our fits to the 
dust temperatures. 
This may be an important source of error.  For example, for HD107146, where 
the SED is particularly well constrained, the best fit SED implies 
$T=51$ K and $\beta=0.7$ (Williams et al.\ 2004), 
compared to a temperature of 43 K that is 
obtained assuming $\beta=1$.  This $\sim 20$\% difference in temperature 
translates into a factor of 40\% error in dust radius.  This is a 
significant difference although much smaller than the dispersion in Figure 5.  

Finally, one might ask how the inferred dust radii compare with 
the spatially resolved sizes of debris disks where available. 
The radial extent of the excess emission, when spatially resolved at 
long wavelengths, generally agrees within a factor of 1.5 with 
the dust radii estimated from the SED. 
Examples of such systems include Fomalhaut (Holland et al.\ 1998), 
HR4796 (Jayawardhana et al.\ 1998; Koerner et al.\ 1998), 
and Vega (Wilner et al.\ 2002).  
In contrast, the sizes of debris disks observed at short wavelengths 
(optical through near-infrared) are typically larger than the sizes 
inferred for the same systems based either on their SEDs at longer 
wavelengths or through resolved millimeter or submillimeter 
imaging.  Examples of such systems include 
Vega (Su et al.\ 2005), 
$\beta$ Pic (Holland et al.\ 1998; Kalas \& Jewitt 1995) and 
AU Mic (Liu et al.\ 2004; Kalas et al.\ 2004). 
In such cases, the short wavelength scattered light probably 
traces a small grain population that is either being ejected 
or placed into eccentric orbits 
through a combination of radiation pressure, corpuscular 
(stellar wind) drag, 
and dynamical scattering with massive planets 
(e.g., Moro-Martin \& Malhotra 2002, 2003b). 

Thus, the available measurements suggest that the dust radii determined 
from the SED assuming blackbody grains reasonably estimates  
the radius of the emitting grains. 
This could be confirmed by spatially resolving the submillimeter 
excess emission in a larger number of sources. 
When such measurements are available it will be interesting to 
examine the true behavior of dust radius as a function of the age 
of the system. 
At the moment, it appears that there is significant dispersion in 
dust radii as a function of age, with no strong correlation between 
these two quantities.

\section{Diverse Evolutionary Histories?}

In summary, the dust masses of the 850 \micron\ disks 
are roughly consistent with the predictions of recent models 
for the collisional evolution of planetesimal disks, in that the mass 
declines with time in the manner expected.  However, the radii of the 
disks suggest a more diverse set of evolutionary histories for 
these systems that extend beyond histories dominated by self-stirring. 

Dominik \& Decin (2003) similarly recognized the need for diversity 
in evolutionary histories in order to explain the large dispersion 
in the fractional luminosities ($f_d = L_{IR}/L_*$) observed by ISO 
in stars spanning a range of ages.  In particular, in order to explain 
the large fractional luminosities ($L_{IR}/L_* = 10^{-3}$) observed by 
ISO in Vega-like systems both young ($\sim 10$ Myr) and old 
($\gtrsim 1$ Gyr), they invoked the idea of ``delayed stirring'' in 
which a primoridal planetesimal disk is allowed to collide only after 
a specified waiting period.  
They speculated that since the Kenyon and 
Bromley models allow for the possibility of dust production over 
a long timescale, as protoplanet formation 
progresses from small to large radii, 
the models may describe the underlying physical mechanism that 
produces luminous dust disks spanning a wide range of ages. 
The distribution of dust disk radii as a function of age for the 
850\micron\ disks sample suggests that this is not the whole story. 

More specifically, some of the properties of individual systems are 
difficult to explain in detail 
solely with the self-stirring model.  
For example, HD104860 has the most massive ($0.16 M_\oplus$) 
as well as the largest dust disk.  
Compared to the $\sim 40-50$ AU disk sizes deduced for HD107146 and HD8907,
the $\sim 33$ K dust temperature of HD104860 corresponds to
a much larger orbital distance of $\sim 110$ AU.
While the large disk size of HD104860 may reflect the more rapid collisional
evolution (and inner disk clearing) expected for more massive disks,
the large size cannot be accounted for with 
a simple scaling of the Kenyon \& Bromley results.  
In their model, it would take $\sim 400$ Myr for the wave of 
collisional evolution to reach 100 AU even for a system with a 
mass 10 times the mass of the minimum mass solar nebula, an 
order of magnitude larger than the published age of 40 Myr for 
the system. 

These results suggest that processes beyond self-stirring play 
a significant role in the evolutionary histories of planetesimal disks. 
One source
of heterogeneity may be the influence of giant planets on planetesimal
disks.  Giant planets may clear regions of planetesimal disks and 
pump up the eccentricities and inclinations of planetesimals in other 
regions of disks, as well as sculpt the resulting dust debris. 
These effects have been invoked to explain the azimuthal asymmetries 
observed in the dust distributions surrounding some of the 
sources in the 850 \micron\ disks sample (e.g., Vega; Wilner et al.\ 2002). 
It would not be surprising if giant planets were present in a 
large fraction of debris disk systems given the current detection 
statistics of precision radial velocity searches 
for extra-solar planets and the likelihood that more giant planets 
remain to be discovered at larger orbital radii (Marcy et al.\ 2004). 

Giant planet formation is of course believed to have had a significant 
impact on the evolutionary history of our own Kuiper Belt. 
The fact that the dust content of the Kuiper Belt in
the solar system is likely located far below the points in Figure 4 
at $\sim 10^{-5} M_\oplus$ (Moro-Martin \& Malhotra 2003a; 
Backman, Dasgupta \& Stencel 1995)
at an age of $\sim 5$ Gyr, is likely the result of a more complex
evolutionary history for the Kuiper Belt than the collisional
grinding of a self-stirred planetesimal disk. 
Indeed, the low density of Kuiper Belt objects in the
solar system today and the resulting low current rate of dust 
production is believed to be the consequence of giant planet
formation which significantly depleted the planetesimal disk in the
30--50 AU region in the early solar system (e.g., Hahn \& Malhotra 1999).

Since processes such as giant planet formation may produce 
significant dispersion about 
the evolutionary trend predicted by self-stirred models,   
large samples of debris disk systems 
spanning a large range in age  
may be needed to recognize the 
expected trend of dust radius increasing with age. 
This may be an interesting topic to return to when larger samples  
are available.

\section{Conclusions}
To conclude, we return to the question raised at the beginning of
section 4.3, on the applicability of self-stirred models to the debris
disks detected in our sample.  As discussed above, there is some
question as to whether self-stirred models apply.  The sizes and
ages of the dust disks surrounding HD8907 and HD107146 appear to
fit the picture described by Kenyon \& Bromley (2004).  If the model 
is applicable to these systems, it suggests that the debris in these 
systems arises from massive planetesimal disks with initial masses 
several times the minimum mass solar nebula.  
In contrast, the
radius of the dust disk surrounding HD104860 appears too large to
explain by a simple scaling of the model.  More generally, the
properties of the 850\micron\ disks sample as a whole suggest the
role of additional processes, such as giant planet formation, in
contributing to the production and shaping of the debris that is
observed in at least some of the systems.  

On the one hand, this is encouraging for the FEPS program, 
since one of the aims of the program is to deduce the underlying 
giant planetary architectures of young solar analogues from the 
properties of their associated debris disks. 
On the other hand, the possibility that multiple processes 
affect the appearance of debris disks suggests that an important 
step toward the FEPS objective is to find ways to distinguish the 
relative roles of potentially multiple processes (e.g., embedded 
protoplanets, inner giant planets) in producing and shaping the debris. 
Spatially resolved observations of the debris may be one such 
discriminant.

\acknowledgments
Our study made use of the significant work by the FEPS team that went 
into compiling and characterizing the FEPS sources.  We would like to 
thank Steve Strom and Michael Liu for careful readings of the 
manuscript. JPW acknowledges support from NSF grant AST-0324328.
This research has made use of the SIMBAD database and
the Two Micron All Sky Survey, which is a joint project of the
University of Massachusetts and IPAC/Caltech, funded by NASA and NSF.

\clearpage
\section{References}
\parskip=0pt
\bigskip

\pp Aikawa, Y., Miyama, S. M., Nakano, T., \& Umebayashi, T. 1996, 
	ApJ, 467, 684

\pp Allen, R. L., Bernstein, G. M., \& Malhotra, R. 2001, ApJ, 549, L241

\pp Ardila, D. R., Golimowski, D. A., Krist J. E., Clampin, M.,
	Williams, J. P., Blakeslee, J. P., Ford, H. C., Hartig G. F.,
	\& G. D. Illingworth, G. D. 2004, ApJ, 617, L147

\pp Aumann, H. H. 1985, PASP, 97, 885

\pp Backman, D. E., Dasgupta, A., \& Stencel, R. E., 1995, ApJ, 450, L35

\pp Backman, D. E., \& Paresce, F. 1993, in 
	Protostars and Planets III, 
	ed. Levy, E. H., \& Lunine, J. I. 
	(Tucson: University of Arizona Press), p. 1253

\pp Beckwith, S. V. W., Henning, T., \& Nakagawa, Y.  2000, in 
	Protostars and Planets IV, 
	ed. Mannings, V., Boss, A. P., \& Russell, S. S. 
	(Tucson: University of Arizona Press), p. 533

\pp Calvet, N., D'Alessio, P., Hartmann, L., Wilner, D., Walsh, A., 
	\& Sitko, M. 2002, ApJ, 568, 1008

\pp Carpenter, J., Wolf, S., Schreyer, K., Launhardt, R., \& Henning, T. 2005,
	AJ, 129, 1049

\pp Chiang, E. I., \& Brown, M. E. 1999, AJ, 118, 1411 

\pp D'Alessio, P., et al.\ 2005, ApJ, 621, 461

\pp Dent, W. R. F., Walker, H. J., Holland, W. S., \& Greaves, J. S. 
	2000, MNRAS, 314, 702 

\pp Dent, W. R. F., Greaves, J. S., Mannings, V., Coulson, I. M., 
	\& Walther, D. M. 1995, MNRAS, 277, L25
 
\pp Dominik, C., \& Decin, G. 2003, ApJ, 598, 626

\pp Farinella, P. Davis, D. R., \& Stern, S. A. 2000, in 
	Protostars and Planets IV, 
	ed. Mannings, V., Boss, A. P., \& Russell, S. S. 
	(Tucson: University of Arizona Press), p. 1255

\pp Feigelson, E. D. \& Nelson, P. I. 1985, ApJ, 293, 192

\pp Gorti, U. \& Hollenbach, D. 2004, ApJ, 613, 424

\pp Greaves, J. S. et al. 1998, ApJ, 506, L133

\pp Greaves, J. S., Coulson, I. M., \& Holland, W. S. 2000, 
	MNRAS, 312, L1

\pp Greaves, J. S., Wyatt, M. C., Holland, W. S., \& Dent, W. R. F. 
	2004a, MNRAS, 351, L54

\pp Greaves, J. S., Holland, W. S., Jayawardhana, R., Wyatt, M. C., 
	\& Dent, W. R. F. 2004b, MNRAS, 348, 1097

\pp Habing, H. et al.\ 2001, A\&A, 365, 545

\pp Hillenbrand, L. A., et al. 2005, submitted

\pp Holland, W. S. et al. 1998, Nature, 392, 788

\pp Hollenbach, D., Yorke, H. W., \& Johnstone, D. 2000, in 
	Protostars and Planets IV, 
	ed. Mannings, V., Boss, A. P., \& Russell, S. S. 
	(Tucson: University of Arizona Press), p. 401

\pp Jayawardhana, R., Fisher, S., Hartmann, L., Telesco, C., 
	Pi\~na, R., \& Fazio, G. 1998, ApJ, 503, L79

\pp Jura, M., Malkan, M., White, R., Telesco, C., Pi\~na, R., \& 
	Fisher, R. S., 1998, ApJ, 505, 897 

\pp Jura, M. et al.\ 2004, ApJS, 154, 453

\pp Kalas, P., Liu, M. C., \& Matthews, B. C. 2004, Science, 303, 1990

\pp Kalas, P., \& Jewitt, D. 1995, AJ, 110, 794

\pp Kamp, I. \& Bertoldi, F. 2000, A\&A, 353, 276

\pp Kastner, J. H., Zuckerman, B., Weintraub, D. A., \& Forveille, T. 
	1997, Science, 277, 67

\pp Kenyon, S. J. \& Bromley, B. C. 2004a, AJ, 127, 513 

\pp Kenyon, S. J. \& Bromley, B. C. 2004b, ApJ, 602, L133

\pp Kenyon, S. J. \& Luu, J. X.  1998, AJ, 115, 2136

\pp Koerner, D. W., Ressler, M. E., Werner, M. W., \& Backman, D. E. 
	1998, ApJ, 503, L83

\pp Krivov, A. V., Mann, I., \& Krivova, N. A., 2000, A\&A, 362, 1127

\pp Lavalley, M., Isobe, T., \& Feigelson, E. 1992,
	in ASP Conf. Ser. 25, Astronomical Data Analysis Software and 
	Systems I, ed. D. M. Worrall, C. Biemesderfer, \& J. Barnes 
	(San Francisco: ASP), 245

\pp Levison, H. F., \& Morbidelli, A. 2003, Nature, 426, 419

\pp Liou, J.-C. \& Zook, H. A. 1999, AJ, 118, 580 

\pp Liu, M. C., Matthews, B. C., Williams, J. P., \& Kalas, P. G.
    2004, ApJ, 608, 526

\pp Marcy, G.~W., \& Butler, R.~P.\ 1998, \araa, 36, 57 

\pp Marcy, G. W., Butler, R. P., Fischer, D. A., \& Vogt, S. S.  2004, 
	ASP Conf. Ser. vol. 321, p. 3

\pp Meyer, M., et al. 2004, ApJS, 154, 422

\pp Miyake, K., \& Nakagawa, Y. 1993, Icarus, 106, 20

\pp Moro-Mart\'in, A. \& Malhotra, R. 2002, AJ, 124, 2305 

\pp Moro-Mart\'in, A. \& Malhotra, R. 2003a, AJ, 125, 2255

\pp Moro-Mart\'in, A. \& Malhotra, R. 2003b, BAAS, 203, 1711

\pp Ozernoy, L. M., Gorkavyi, N. N., Mather, J. C., \& Taidakova, T. A. 
	2000, ApJ, 537, L147 

\pp Pollack, J. B., Hollenbach, D., Beckwith, S., Simonelli, D. P., 
	Roush, T., \& Fong, W.  1994, ApJ, 421, 615

\pp Qi et al. 2004, ApJ, 616, L11

\pp Richter, M. J., Jaffe, D. T., Blake, G. A., \& Lacy, J. H. 2002, 
	ApJ, 572, L161

\pp Quillen, A. C., \& Thorndike, S. 2002, ApJ, 578, L149 

\pp Sako, S. et al.\ 2005, ApJ, 620, 347

\pp Sandford, S. A., \& Allamandola, L. J. 1990, Icarus, 87, 188

\pp Sandford, S. A., \& Allamandola, L. J. 1988, Icarus, 76, 201

\pp Scoville, N. Z., Sargent, A. I., Sanders, D. B., Claussen, M. J., 
    Masson, C. R., Lo, K. Y., \& Phillips, T. G. 1986, ApJ, 303, 416

\pp Sheret, I., Dent, W. R. F., Wyatt, M. C. 2004, MNRAS, 348, 1282

\pp Sheret, I., Ramsay Howat, S. K., \& Dent, W. R. F. 2003, MNRAS, 343, L65

\pp Spangler, C., Sargent, A. I., Silverstone, M. D., Becklin, E. E.,
    \& Zuckerman, B. 2001, ApJ, 555, 932

\pp Silverstone, M. D. 2000, Ph. D. Thesis, UCLA

\pp Stern, S. A. 1995, AJ, 110, 856

\pp Stern, S. A. \& Colwell, J. E., 1997, AJ, 114, 841

\pp Su, K. Y. L. et al.\ 2005, astro-ph/0504086

\pp Sylvester, R. J., Dunkin, S. K., \& Barlow, M. J. 2001, 
	MNRAS, 327, 133

\pp Takeuchi, T., \& Artymowicz, P. 2001, ApJ, 557, 990 

\pp Thommes, E. W., Duncan, M. J., \& Levison, H. F. 2003, Icarus, 161, 431

\pp Tielens, A. G. G. M., \& Allamandola, L. J. 1987, in 
	"Interstellar Processes", ed. D. J. Hollenbach \& 
	H. A. Thronson, Jr. (Dordrecht: Reidel), p. 397

\pp Thi, W. F., et al.\ 2001, ApJ, 561, 1074

\pp van Dishoeck, E. F., \& Black, J. H. 1988, ApJ, 334, 771

\pp Williams, J. P., Najita, J., Liu, M. C., Bottinelli, S.,
    Carpenter, J. M., Hillenbrand, L. A., Meyer, M. R., \& Soderblom, D. R.
    2004, ApJ, 604, 414 (Paper I)

\pp Wilner, D. J., Holman, M. J., Kuchner, M. J., \& Ho, P. T. P.
    2002, ApJ, 569, L115

\pp Wright, E. L. 1987, ApJ, 320, 818

\pp Wyatt, M. C. 2005, A\&A, 433, 1007

\pp Wyatt, M. C., Greaves, J. S., Dent, W. R. F., \& Coulson I. M. 2005, 
	ApJ, 620, 492

\pp Wyatt, M. C., Dent, W. R. F., \& Greaves, J. S. 2003, MNRAS, 342, 876

\pp Zuckerman, B., \& Becklin 1993

\pp Zuckerman, B., Forveille, T., \& Kastner, J. H., 1995, 
	Nature, 373, 494

\clearpage
\begin{table}
\begin{center}
TABLE 1\\
Summary of JCMT/SCUBA measurements\\
\vskip 2mm
\begin{tabular}{lcccccccccl}
\hline\\[-2mm]
Source & SpT & Age & $d$ & $\lambda$ & $F_\nu^{\rm a}$ & $M_{\rm d}^{\rm b}$ \\
 & & (Myr) & (pc) & ($\mu$m) & (mJy) & ($M_\oplus$) \\[2mm]
\hline\hline\\[-3mm]
%%
%% I have changed the ages and distances of some sources.
%% Please change the masses.
%%
1RXS~J072343 & K3 & 130 & 24   & 850 &  $< 3.9$     & $< 0.014$ \\
HD~17925     & K1 & 100 & 10.4 & 850 &  $< 7.2$     & $< 0.005$ \\
V383~Lac     & K0 &  60 & 50   & 850 &  $< 8.1$     & $< 0.12$  \\
HD~104860    & F8 &  40 & 47.9 & 850 & $6.8\pm 1.2$ & $0.16$    \\
             &    &     &      & 450 & $47\pm 11$   &           \\
HD~8907      & F8 & 180 & 34.2 & 850 & $4.8\pm 1.2$ & $0.036$   \\
             &    &     &      & 450 & $22\pm 11$   &           \\
HD~35850     & F7 &  12 & 26.8 & 850 &  $< 5.4$     & $< 0.024$ \\
HD~984       & F5 &  40 & 46.2 & 850 &  $< 4.5$     & $< 0.059$ \\
SAO~150676   & F5 &  60 & 78   & 850 &  $< 6.3$     & $< 0.23$  \\
HD~88638$^c$ & G5 &  10 & 37.5 & 850 &  $< 4.8$     & $< 0.041$ \\
HD~217343    & G3 &  40 & 32.0 & 850 &  $< 6.0$     & $< 0.037$ \\
HD~107146    & G2 & 100 & 28.5 & 850 & $20\pm 3.2$  & $0.10$    \\
             &    &     &      & 450 & $130\pm 12$  &           \\
HD~166435$^c$& G0 &  10 & 25.2 & 850 &  $< 6.9$     & $< 0.027$ \\
HD~77407     & G0 &  30 & 30.1 & 850 &  $< 5.1$     & $< 0.028$ \\[2mm]
\hline\\[-3mm]
\multicolumn{7}{l}{$^{\rm a}$ Upper limits are $3\sigma$, errors on the
detections are $\pm\sigma$}\\
\multicolumn{7}{l}{$^{\rm b}$ $\kappa=1.7$~cm$^2$~g\e;
$T=50$~K for upper limits, 33~K for HD~104860,}\\
\multicolumn{7}{l}{48~K for HD~8907, and 51~K for HD~107146 }\\
\multicolumn{7}{l}{$^{\rm c}$ May be significantly older; dropped from 
FEPS sample subsequent}\\ 
\multicolumn{7}{l}{to selection for this study.}
\end{tabular}
\end{center}
\end{table}

\clearpage
\begin{table}
\begin{center}
TABLE 2\\
$\beta=1$ SED fits\\
\vskip 2mm
\begin{tabular}{llccc}
\hline\\[-2mm]
Source & & $T_{dust}$ & $M_{dust}$  &  $L_{dust}$ \\
       & &  (K)  & ($M_\oplus$) & ($10^{-5}L_\odot$) \\[2mm]
\hline\hline\\[-3mm]
HD~104860 & best & 33 &  0.16  &  46 \\
          & min  & 19 &  0.34  &   8 \\
          & max  & 42 &  0.12  &  80 \\
HD~8907   &      & 48 &  0.036 &  54 \\[2mm]
\hline
\end{tabular}
\end{center}
\end{table}

\clearpage
\begin{figure}[ht]
\plotfiddle{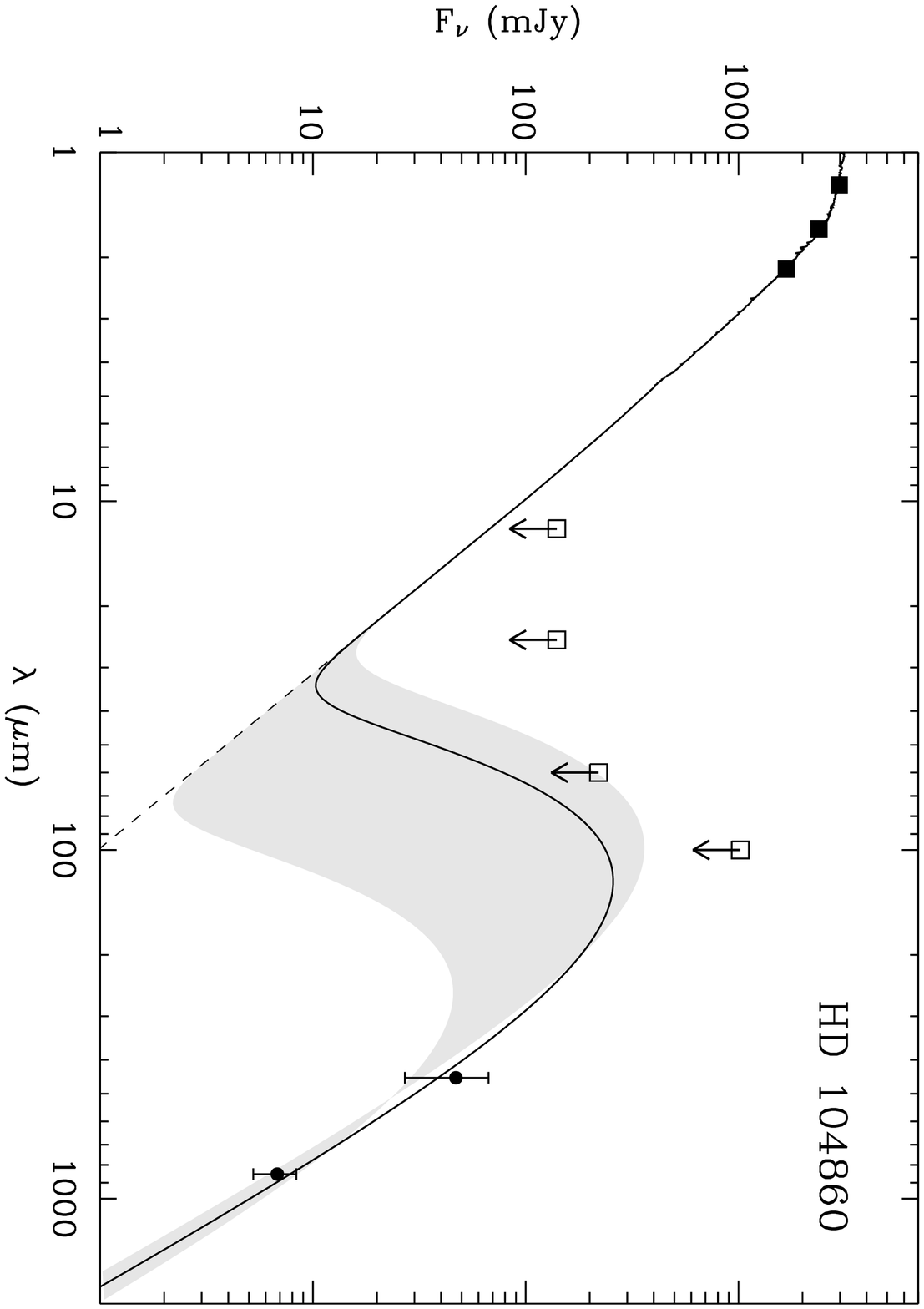}{50pt}{90}{60}{60}{230}{-280}
\end{figure}
\vskip 3.4in
\noindent{\bf Figure 1:}
Spectral energy distribution of HD~104860.
The three filled squares are from the 2MASS catalog.
The star-disk system was not detected by IRAS; the open
squares represent the FSC upper limits color-corrected for
a dust temperature of 6000~K at 12 and 25~\micron\ and
40~K at 60 and 100~\micron.
The filled circles show the SCUBA measurements at
450~\micron\ and 850~\micron\ with $\pm\sigma$ error
bars. The solid line shows the best fit to the stellar photosphere
and disk emission ($T_{dust}=33$~K, $\beta=1$), the dashed line shows the
extrapolated contribution from the photosphere in the
far-infrared. The grey area shows possible fits that do
not violate the IRAS upper limits and lie within the systematic
and calibration uncertainties of the SCUBA measurements.
The corresponding range in dust mass and temperature 
is given in Table 2.

\clearpage
\begin{figure}[ht]
\plotfiddle{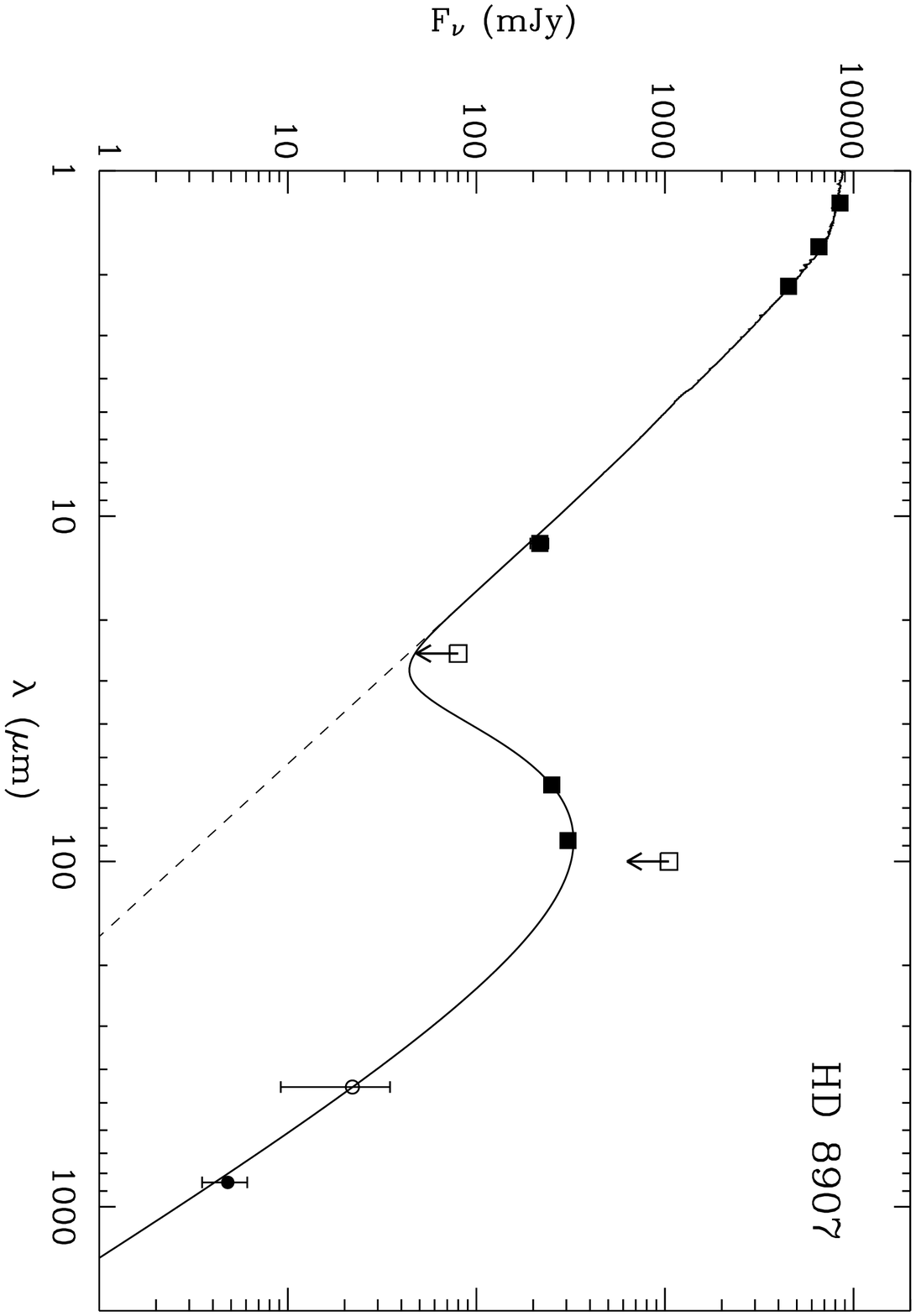}{50pt}{90}{60}{60}{230}{-280}
\end{figure}
\vskip 3.4in
\noindent{\bf Figure 2:}
Spectral energy distribution of HD~8907.
Detections are shown as filled symbols and include
2MASS photometry in the near-infrared, the 12~\micron\
measurement from the IRAS FSC,
60 and 87~\micron\ measurements from ISO,
and the 850~\micron\ SCUBA measurement.
The $2\sigma$ SCUBA detection at 450~\micron\ is shown as an
open circle. For all points the error bars are $\pm\sigma$
(when larger than the symbol size).
The solid line shows the best fit to the stellar photosphere
and disk emission ($T_{dust}=48$~K, $\beta=1$).

\clearpage
\begin{figure}[ht]
\plotfiddle{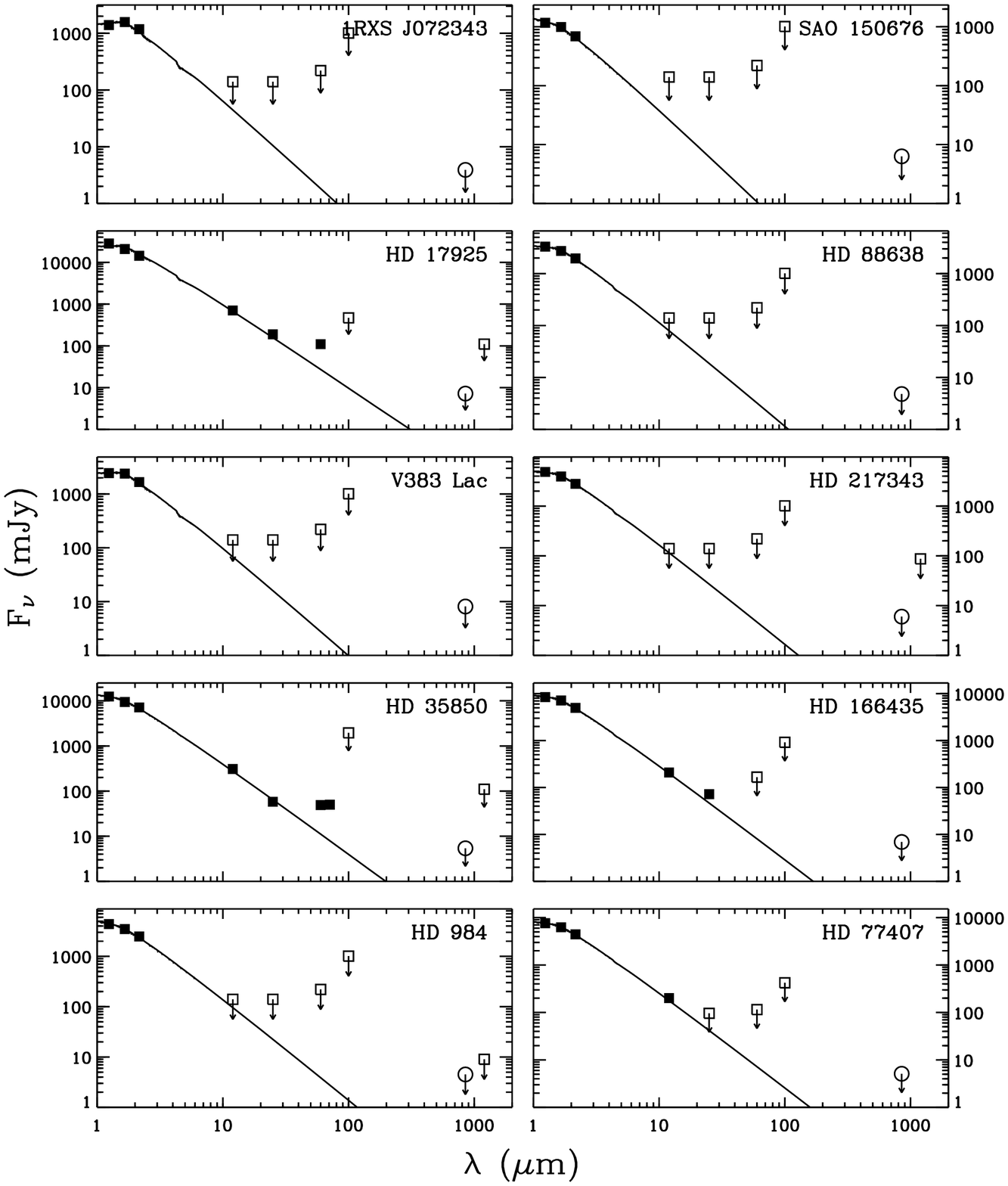}{50pt}{0}{70}{70}{-220}{-440}
\end{figure}
\vskip 5.6in
\noindent{\bf Figure 3:}
Spectral energy distributions for the 10 sources in the sample
that were not detected in the submillimeter. The cluster of 3 points
at $1-2$~\micron\ are fluxes from the 2MASS catalog which were used
to normalize Kurucz models appropriate for each stellar type.
IRAS and ISO measurements from $12-100$~\micron\ are shown as
solid squares for detections and open symbols with an arrow for
upper limits. A few stars were observed at 1.3~mm by
Carpenter et al. (2005), and these non-detections are
also shown as open squares.
The $3\sigma$ SCUBA 850~\micron\ upper limits from this work
are shown as open circles.

\clearpage
\begin{figure}[ht]
\plotfiddle{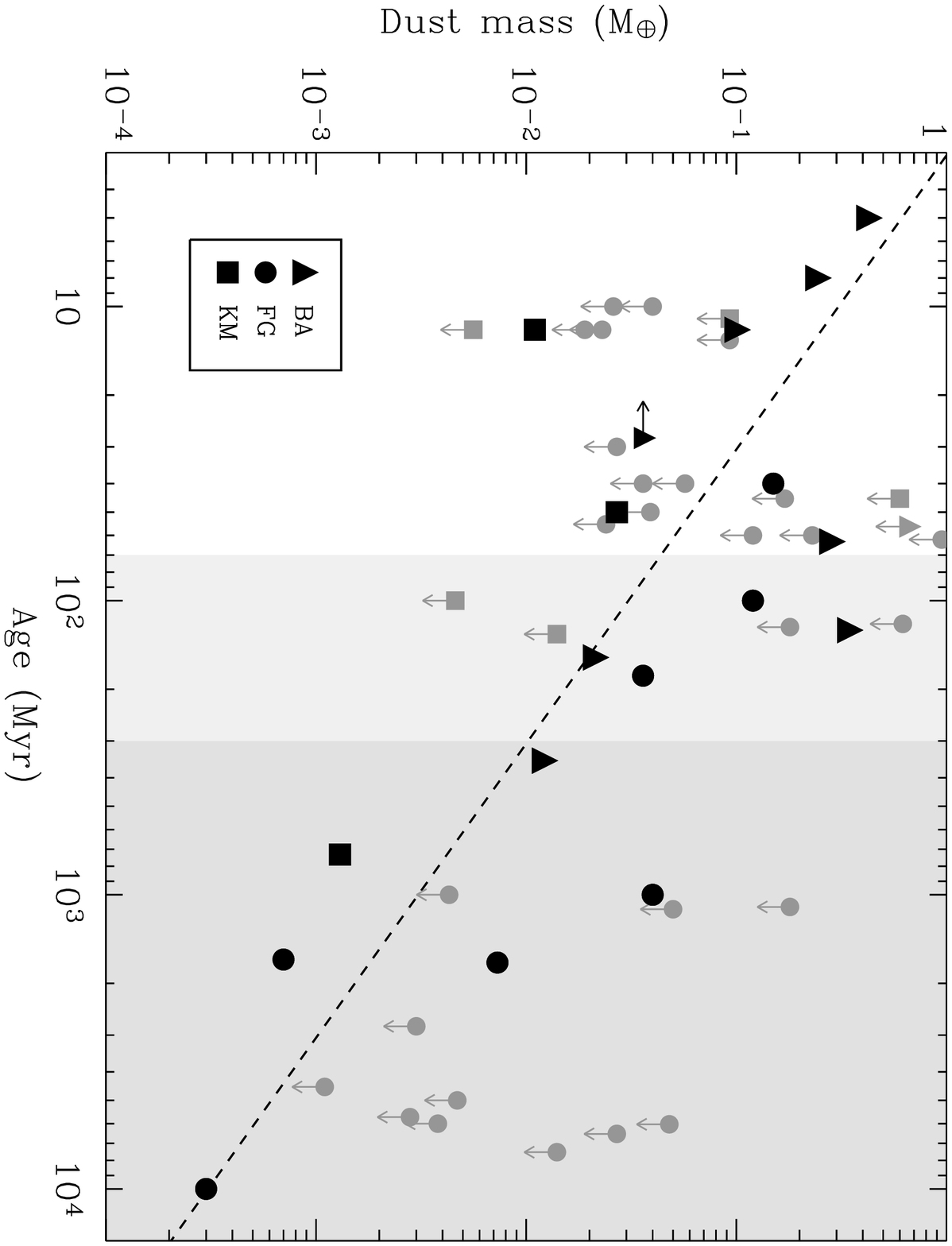}{50pt}{90}{60}{60}{230}{-280}
\end{figure}
\vskip 3.6in
\noindent{\bf Figure 4:}
Disk masses versus estimated stellar age for the sample of 13 objects
discussed in this paper and for other sources in the literature.
Detected sources are shown in black and grey points signify
$3\sigma$ upper limits based on a 50~K disk temperature, the average
temperature for disks detected around solar type stars.
The different symbols represent different stellar spectral types,
as shown in the insert.
The dashed line shows the slope expected if dust mass declines 
inversely with age and the darker shaded area shows the relatively
older population in contrast to a significantly younger population
in the unshaded region. The 70-300~Myr border between the two,
due to low number statistics and age uncertainties, is shown
by the lighter shaded vertical stripe.

\clearpage
\begin{figure}[ht]
\plotfiddle{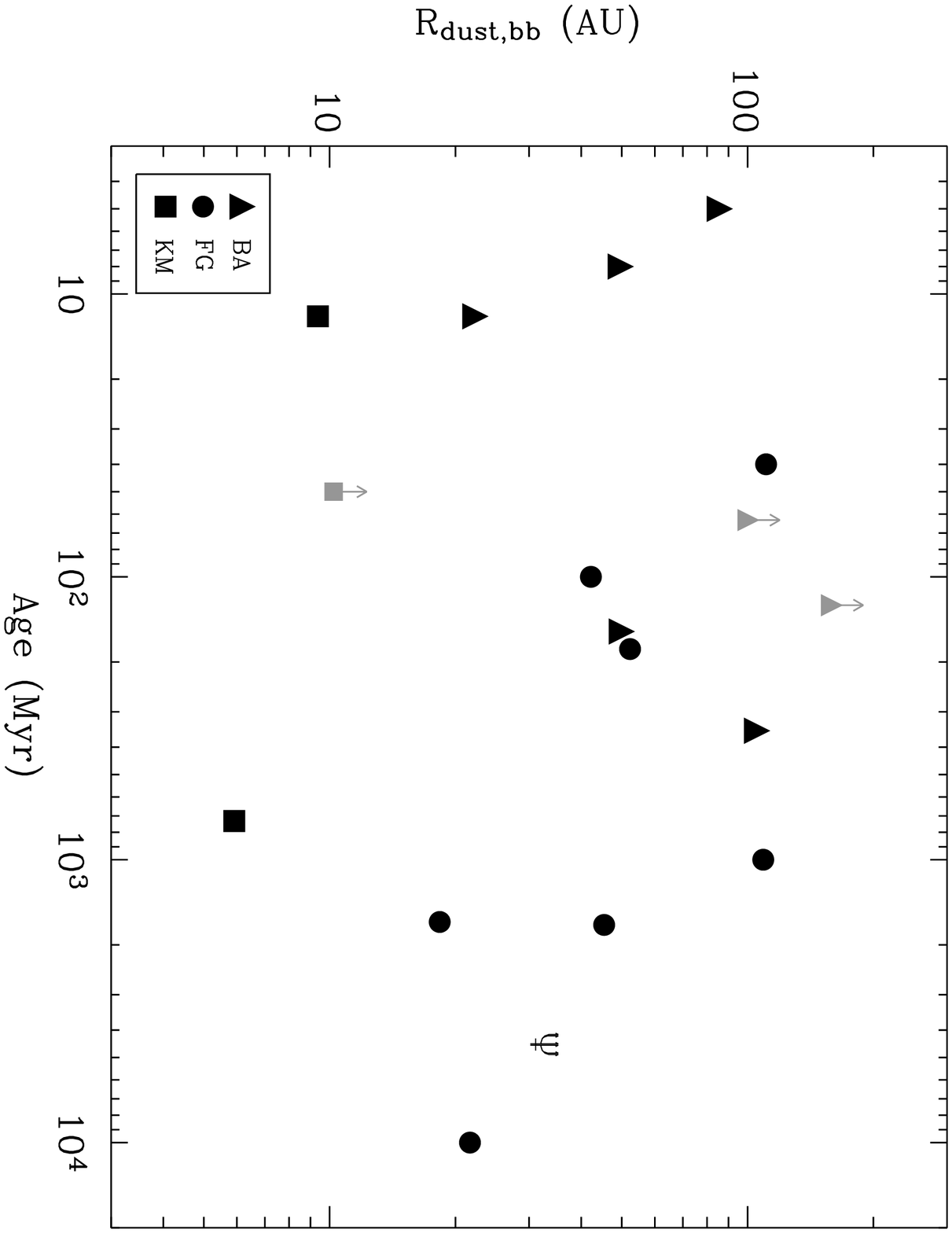}{50pt}{90}{60}{60}{240}{-280}
\end{figure}
\vskip 3.6in
\noindent{\bf Figure 5:}
Characteristic dust emission radii, assuming blackbody grains, 
plotted against estimated stellar age for the three detections 
in this paper and other sub-millimeter detections in the literature. 
The different symbols represent different stellar spectral types,
as described in Figure 4.
The upper limits represent the submillimeter-only excess sources from 
Wyatt et al.\ (2003). 
The trident symbol indicates the orbit of Neptune in the solar system.

\end{document}